# Ultrafast coherent energy transfer with high efficiency based on plasmonic nanostructures

Jun Ren, Tian Chen, Bo Wang and Xiangdong Zhang*
Beijing Key Laboratory of Nanophotonics & Ultrafine Optoelectronic Systems, School of Physics, Beijing Institute of Technology, 100081, Beijing, China.
** Corresponding author, E-Mail address:zhangxd@bit.edu.cn*

## ABSTRACT

The theory of energy transfer dynamics of a pair of donor and acceptor molecules located in the plasmonic hot spots is developed by means of the master equation approach and the electromagnetic Green's tensor technique. A nonlocal effect has been considered by using a hydrodynamic model. The coherent interaction between the two molecules in plasmonic nanostructures is investigated under some strong coupling conditions. It is shown that the energy transfer efficiency of a pair of molecules can be improved largely and the transfer time decreases to dozens of femtoseconds when the contribution of quantum coherence is considered. The physical origin for such a phenomenon has also been analyzed. This ultrafast and high-efficiency energy transfer mechanism could be beneficial for artificial light-harvesting devices.

## I. INTRODUCTION

Energy transfer is an important topic in the world. Many processes in nature involve energy transfer. One of the most famous is the photosynthesis in green plants, in which the energy transfer efficiency can reach nearly 100%[1,2]. Inspired by the photosynthesis, many researches have focused on the efficient energy transfer in artificial structures, and derived many applications. For example, efficient energy transfer mediated by plasmonic nanostructure can be used in solar energy conversion[3], sensitive and efficient biosensing has been realized by designing the novel nanostructures[4], and white-light emitting nanofiber has been fabricated through the efficient energy transfer[5]. There exists several different theories to describe the energy transfer between the donor molecule and acceptor molecule. The semiclassical Förster theory[6-8] has been put forward to interpret the energy transfer when there is an overlap between the emission spectra of donor and absorption spectra of acceptor. This theory has achieved much success in describing the energy transfer when the separation distance between the donor and acceptor is in the sub-wavelength range, and has many important applications like fluorescence resonance energy transfer technique. Another theory is the Dexter theory[9], which has been demonstrated to work well when the separation distance between the donor and the acceptor is less than 3 nm. However, the ultrafast and high-efficiency energy transfer in the photosynthesis and some nanostructures is not still understood very well. An important problem is that quantum coherence in above theories has been neglected.

Recent investigations have shown that quantum coherence may play an important role in the energy transfer process of photosynthesis[10-18]. The quantum coherence between different energy transfer pathways could enhance the efficiency of energy transfer[19-26]. Such a large coherence is not exclusive to the photosynthesis, it also appears in other systems. For example, in many strong coupling nanostructures like optical nanocavity[27,28] and photonic-crystal microcavity[29], it has been demonstrated that the coherence couplings between two quantum emitters could be large with the help of resonance laser. Moreover, in the donor-bridge-acceptor model study[30], it has been confirmed that quantum coherence effect can play an important role in the dynamics of the energy transfer. The stronger the coupling between the donor and acceptor, the greater the effect of the coherence effect on the energy transfer. However, it is difficult to construct very strong coupling between the donor and acceptor in usual nanostructures[31-34].

Very recently, it is reported that the strong resonant coupling between two molecules has been realized in the hot spots of plasmonic systems[35], and simultaneous large energy transfer rate and efficiency has been obtained. However, quantum coherence effects have not been considered in such a case. The problem is how the energy transfer process in such a case will be affected by the quantum coherence effect? In this paper, we study the energy transfer between the two molecules (donor and acceptor) located in the plasmonic system composed of three coaxial nanoparticles. Due to the contribution from the three nanopartilces, the coherent coupling between the donor and the acceptor can be increased greatly. By employing the master equation method, we develop a theory that can be used in investigating how the quantum coherence between the molecules affects the energy transfer. Compared with the results without taking the coherence into account, the emergence of quantum coherence can help to increase the efficiency of the energy transfer and decrease the transfer time. Moreover, the nonlocal effect from the nanoparticles on the efficiency of the energy transfer have been studied.

The rest of this paper is arranged as follows: in Sec. II, we present the description of our system, and then give the coherent energy transfer theory based on the plasmonic hot spots. The detailed calculations and discussions are presented in Sec. III. The nonlocal effect on coherent energy transfer is discussed in Sec. IV. A summary is given in Sec. V.

## II. COHERENT ENERGY TRANSFER THEORY BASED ON PLASMONIC HOT SPOTS

We investigate the energy transfer dynamics of a pair of donor and acceptor molecules located in the plasmonic hot spots through the master equation approach. The relevant parameters are calculated by using the electromagnetic Green's tensor technique. The schematic representation of our plasmonic energy transfer system has been addressed in Fig. 1(a), where a couple of donor and acceptor molecules are inserted in the gaps of three coaxial and equally spaced nanoparticles (trimer). The two gaps are often called plasmonic hot spots if the distance between the spheres is very small, like several nanometers.[35,36] The donor and acceptor are marked by D and A, and their position vectors are denoted by $\vec{r}_D$ and $\vec{r}_A$ respectively.

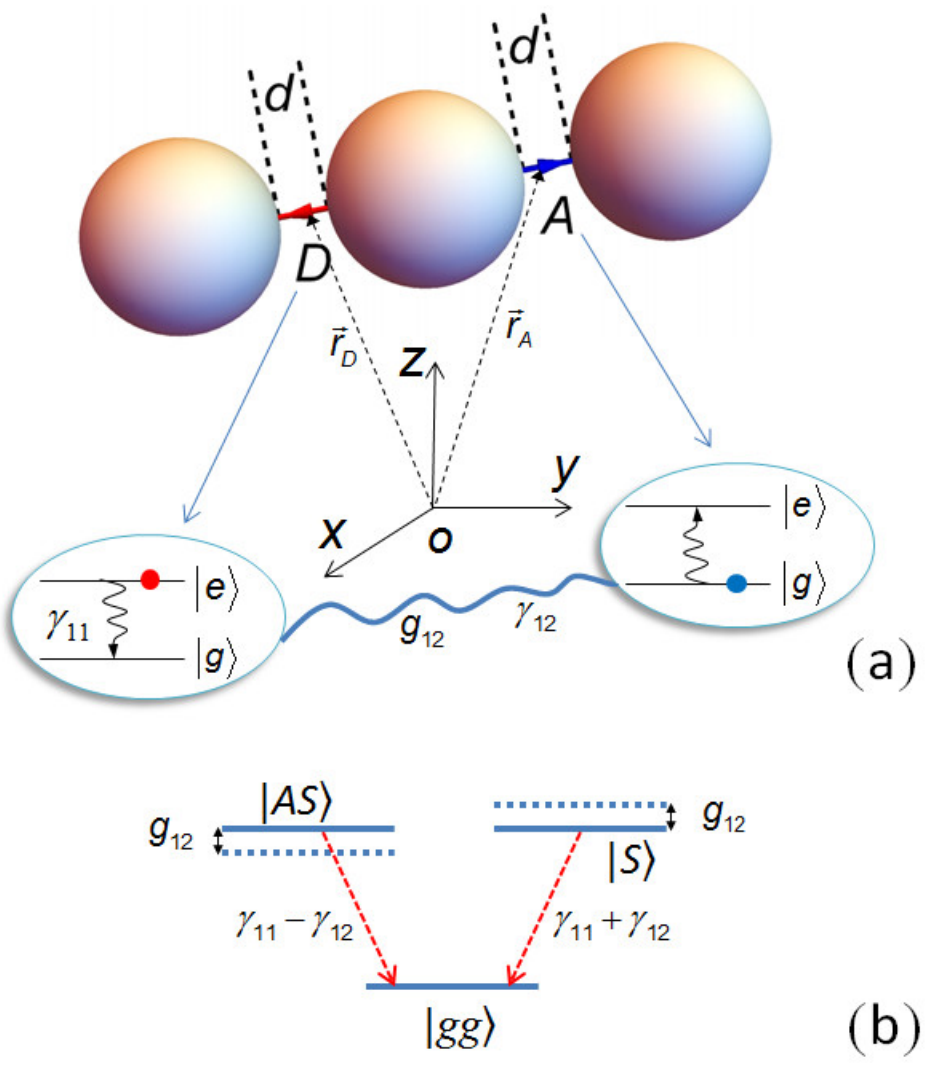

FIG. 1. (a) Energy transfer pattern of a pair of donor and acceptor molecules in plasmonic environment. Initially the excitation is in the donor, and acceptor is in the ground state. $g_{12}$ and $\gamma_{12}$ represent coherent and incoherent terms between the two molecules mediated by environment. (b) Energy-level diagram of eigenstates. $|gg\rangle$ is the ground state, $|S\rangle$ and $|AS\rangle$ are symmetric and anti-symmetric states, $\gamma_{11} \pm \gamma_{12}$ are their spontaneous decay rates, $g_{12}$ is the energy level shift.

We assume that the donor is in excited state and the acceptor is in ground state initially (see the insets of Fig. 1(a), donor is in excited state $|e\rangle$ and acceptor in ground state $|g\rangle$). For the excitation energy in the donor, there are several diffusion approaches: dissipation, radiation into free space or charge separation after the energy migration from the donor to the acceptor. In Fig. 1, the parameters $g_{12}$ and $\gamma_{12}$ representing coherent and incoherent couplings, which are mediated by the multi-mode fields in the trimer system. To study the dynamics of molecules, a full quantum theory has been taken, and the environment is considered as the quantized radiation field. The two molecules are considered as the electric point-dipole of two-level systems, and the orientations of dipole moments along the axis of the trimer. This point-dipole approximation is accurate enough for qubits like small molecules used in this paper. For such an open system, the dynamics of the density operator $\rho$ for the two molecules is described by a Lindblad master equation after tracing out over the environment degree of freedom[23,33,37]

$$\begin{aligned}\partial_t \rho = \frac{i}{\hbar}[\rho, H] &+ \frac{(\gamma_D + 2\gamma_0)}{2}\left(2\sigma_D \rho \sigma_D^\dagger - \sigma_D^\dagger \sigma_D \rho - \rho \sigma_D^\dagger \sigma_D\right) \\ &+ \frac{(\gamma_A + 2\kappa)}{2}\left(2\sigma_A \rho \sigma_A^\dagger - \sigma_A^\dagger \sigma_A \rho - \rho \sigma_A^\dagger \sigma_A\right) \\ &+ \frac{\gamma_{12}}{2}\left(2\sigma_D \rho \sigma_A^\dagger - \sigma_D^\dagger \sigma_A \rho - \rho \sigma_D^\dagger \sigma_A\right) \\ &+ \frac{\gamma_{21}}{2}\left(2\sigma_A \rho \sigma_D^\dagger - \sigma_A^\dagger \sigma_D \rho - \rho \sigma_A^\dagger \sigma_D\right)\end{aligned}, \qquad (1)$$

where $\sigma_{D(A)}^{\dagger}$ and $\sigma_{D(A)}$ are the creation and annihilation operators for the donor (acceptor), and $H$ is the Hamiltonian of the molecules

$$H = \sum_{i=D,A} \hbar(\omega_0 + \delta_i)\sigma_i^{\dagger}\sigma_i + g_{12}\left(\sigma_D^{\dagger}\sigma_A + \sigma_D^{\dagger}\sigma_A\right), \tag{2}$$

in which $\omega_0$ is the transition frequency of the molecules in the first term, in this work the transition frequencies of two molecules are considered as the same. $\delta_i$ is the material-induced Lamb shift induced by molecule-field interaction, which has small contribution to the dynamics[25]. The second term in Hamiltonian plays a role of coherent coupling between the donor and acceptor, with

$$g_{ij} = \frac{1}{\pi\varepsilon_0\hbar}\mathcal{P}\int_0^{\infty}\frac{\omega^2\,\mathrm{Im}\left[\vec{\mu}_i^{*}\cdot\vec{G}(\vec{r}_i,\vec{r}_j,\omega)\cdot\vec{\mu}_j\right]}{c^2\left(\omega-\omega_0\right)}d\omega, \tag{3}$$

which arises from the dipole-dipole interaction between the molecules through the field. In equation (3) $\vec{G}(\vec{r}_i,\vec{r}_j,\omega)$ is the classical Green tensor of the system, which is the solution of the tensor equation $[\nabla\times\nabla\times - k_0^2\varepsilon(\vec{r},\omega)]\vec{G}(\vec{r},\vec{r}';\omega) = \vec{\mathrm{I}}\,\delta(\vec{r}-\vec{r}')$. Here, $\varepsilon(\vec{r},\omega)$ is the relative electric permittivity, and in this work the magnetic response has been omitted. The detailed calculations for Green tensor can be found in Appendix A. The symbol $\mathcal{P}$ stands for the principle integral, and $\vec{\mu}_i$ and $\vec{\mu}_j$ are dipole moments of two molecules. In our calculations, we have taken $\left|\vec{\mu}_i\right| = \left|\vec{\mu}_j\right| = \left|e\right|\times\left|r_0\right|$, where $\left|r_0\right| = 1$ Å. In Eq. (1), $\gamma_0$ represents the dissipation in the donor, and $\kappa$ is the charge separation rate. We have chosen $\gamma_0 = 1\ \mathrm{ns}^{-1}$ and $\kappa = 4\ \mathrm{ps}^{-1}$ in our calculations[23]. The parameter $\gamma_{D(A)} = \gamma_{ii}$ is the dissipation of donor (acceptor) in the environment, here we assume that the donor and acceptor have the same decay rate. The parameter $\gamma_{12(21)}$ is referred to as the incoherent interference term between the two molecules that is mediated by the plasmons

$$\gamma_{ij} = \frac{2\omega_0^2}{\varepsilon_0 c^2\hbar}\mathrm{Im}\left[\vec{\mu}_i^{*}\cdot\vec{G}(\vec{r}_i,\vec{r}_j,\omega)\cdot\vec{\mu}_j\right]. \tag{4}$$

It is clear that, if the dissipation rate in the donor and charge separation rate in the acceptor are constant, the incoherent and coherent parameters $\gamma_{12}$ and $g_{12}$ play a major role in the evolutions of the two molecules. The parameter $\gamma_{12}$ showed in Eq. (4) introduces a coupling between the molecules through the vacuum field. In this case, the spontaneous emission of one of the molecules influences the emission rate of the other. The dipole–dipole interaction term $g_{12}$ introduces a coherent coupling between the molecules. Owing to the dipole–dipole interaction, the excitation is coherently transferred back and forth from one molecule to the other. In this work, all the molecules only involve two energy levels.

There exists two different types of molecule-field interactions in the plasmonic system. One is the resonant interaction, which occurs when the transition frequency of the molecule is resonant with the field, and the interaction strength is proportional to the spontaneous decay rate of the molecule and the imaginary part of the medium-mediated Green tensor.[38] The other is off-resonant interaction. There is a shift between the resonance modes and the transition frequency of molecule, and the interaction strength is proportional to the principle integral of Green tensor as shown in Eq. (3). This off-resonant interaction could induce energy shift, that is the Lamb shift induced by the electric field.

According to the analysis above, molecule-field interaction includes resonant and off-resonance interaction, in fact, molecule-molecule interaction is the same. From Eqs. (3) and (4) we can see that, incoherent interaction is resonant coupling, which could alter the spontaneous emission rates of two molecules. While, coherent interaction is off-resonance coupling, which is the origin of the dipole-dipole interaction between two molecules, corresponds to the energy shift. In vacuum, when the two molecules are departed by a few nanometers distance, their dipole-dipole interaction is very small. However, with the existence of the medium, especially in the regime with continuous electromagnetic modes, the off-resonance molecule-field interaction tends to crucial[39]. When the two interactions have been considered in Hamiltonian, the new two eigenstates of two-molecule system could be obtained by diagonalizing the Hamiltonian[37]

$$\begin{aligned} \left|S\right\rangle &= \frac{1}{\sqrt{2}}(\left|e_1,g_2\right\rangle+\left|g_1,e_2\right\rangle) \\ \left|AS\right\rangle &= \frac{1}{\sqrt{2}}(\left|e_1,g_2\right\rangle-\left|g_1,e_2\right\rangle) \end{aligned} . \qquad (5)$$

The state $\left|e_i,g_j\right\rangle$ represents that the molecule $i$ is in the excited state and molecule $j$ is in the ground state. Figure 1(b) shows the energy-level diagram of eigenstates, where $g_{12}$ denotes the energy-level shift of symmetric state $\left|S\right\rangle$ and anti-symmetric state $\left|AS\right\rangle$. Such a shift is induced by the coherent dipole-dipole interaction between the two molecules. $\gamma_{11}\pm\gamma_{12}$ are the spontaneous emission rates of two eigenstates.

In fact, it should be pointed out that in the derivation of Eq. (1), two approximations have been employed. One is rotating-wave approximation, in which the terms that do not meet the conservation of energy have been removed in the interaction Hamiltonian. The other is the Markov approximation, which requires that the parameters $\gamma_{12}$ and $g_{12}$ do not change very rapidly with the frequencies.

The dissipation rate $\gamma_{11}$ and interference term $\gamma_{12}$ showed in Eq. (1) can be easily obtained from the calculations of Green tensor, which can be found in Appendix A. However for the coherent term appeared in Eq. (3), we have to deal with the principle integral in our two-molecule system. According to Ref. 40 and our more detailed Appendix B, the tough principle integral can be avoided by using the Kramers-Kronig relation and the contour integral, and the coherent term can be simplified as

$$\begin{aligned} g_{ij} &= \frac{\omega_0^2}{\varepsilon_0 \hbar c^2} \mathrm{Re}\left[\vec{\mu}_i^* \cdot \vec{\vec{G}}^s(\vec{r}_i, \vec{r}_j, \omega) \cdot \vec{\mu}_j\right] \\ &+ \frac{1}{\pi \varepsilon_0 \hbar} \int_0^\infty \mathrm{Re}\left[\vec{\mu}_i^* \cdot \vec{\vec{G}}^s(\vec{r}_i, \vec{r}_j, i\kappa) \cdot \vec{\mu}_j\right] \frac{\omega_A}{\omega_A^2 + \kappa^2} d\kappa \end{aligned} . \quad (6)$$

To obtain the efficiency of coherent energy transfer, we need to deal with master Eq. (1). As we have assumed, there is only one excitation in the system, thus there are three possible states for two molecules: (1) the donor is excited and the acceptor is in ground state, denoted by $|D\rangle$, this is also the initial state; (2) the acceptor is excited and the donor is in ground state, denoted by $|A\rangle$; (3) the donor and acceptor are both in ground states, in this case the excitation is dissipated or used for charge separation, denoted by $|0\rangle$. We have the relations $\sigma_D^\dagger|0\rangle = |D\rangle$, $\sigma_A^\dagger|0\rangle = |A\rangle$. The expectation values of both sides of master equation (Eq. (1)) on the state $|0\rangle$ can be given by

$$\dot{\rho}_{00}(t) = (\gamma_D + 2\gamma_0)P_D(t) + \gamma_A P_A(t) + 2\kappa P_A(t), \quad (7)$$

where $\dot{\rho}_{00}(t) = \langle 0|\partial_t \rho(t)|0\rangle = \partial_t \langle 0|\rho(t)|0\rangle$ represents the total probability density of excitation diffused by dissipation and using for charge separation. $P_D(t)$ and $P_A(t)$ are population of donor and acceptor respectively, $P_D(t) = \langle D|\rho(t)|D\rangle$ and $P_A(t) = \langle A|\rho(t)|A\rangle$. The first term of the right side of Eq. (7) represents the probability density of dissipation in the donor, the second term is dissipation in the acceptor, and the third term is the probability density of charge separation in the acceptor, denoted by $\omega_{RC}(t) = 2\kappa P_A(t)$, the meaning is the probability of charge separation happens in $(t, t+dt)$ time interval. Considering the excitation can transfer back and forth between donor and acceptor, the calculation method of energy transfer efficiency[41] in the classical energy theory is not applicable any more. In the coherent energy transfer theory, the energy transfer efficiency is defined as the total probability of excitation used for charge separation[23], $\eta = \int_0^\infty \omega_{RC}(t)dt$ , thus the energy transfer efficiency in our master equation approach is

$$\eta = 2\int_0^\infty \kappa P_A(t)dt\,. \tag{8}$$

Substitute the coupling parameters Eqs. (3) and (4) into master equation, we can solve the density matrix numerically, obtain $P_A(t)$ , and calculate the efficiency denoted by Eq. (7). Aside efficiency, in the coherent energy transfer theory, there is another key factor that is energy transfer time, which is defined as the average waiting time before charge separation happens in the acceptor[23]

$$t_f = \frac{1}{\eta}\int_0^\infty t\omega_{RC}(t)dt\,, \tag{9}$$

thus in our master equation approach, energy transfer time is

$$t_f = \int_0^\infty tP_A(t)dt \Big/ \int_0^\infty P_A(t)dt\,. \tag{10}$$

## III. CALCULATIONS AND DISCUSSIONS ON COHERENT ENERGY TRANSFER EFFICIENCY AND TIME

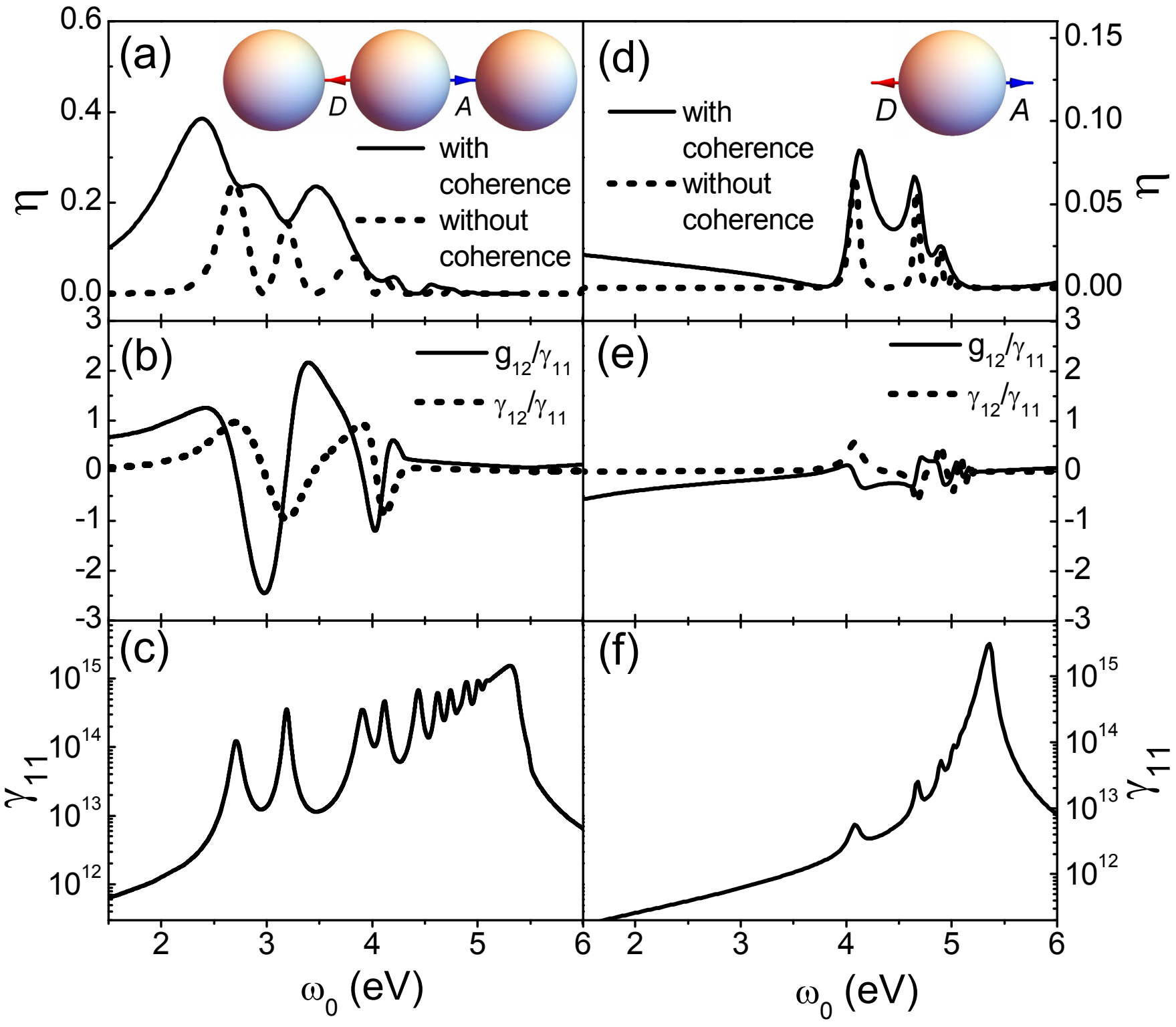

FIG. 2. (a) Black line: Energy transfer efficiency of donor and acceptor as a function of transition frequency located in hot spots of metallic nanosphere trimer showed in Fig. 1. R=10 nm, d=1 nm. Red line: Same for the former case by neglecting the coherence. Blue line: single sphere case. (b) Coherent (solid lines) and incoherent (dashed lines) parameters. (c) Donor decay rates. (d) Dependence of separation distance of the trimer system on the energy transfer efficiency.

We use Eq. (8) to calculate the coherent energy transfer efficiency. Figure 2(a) shows the energy transfer efficiency of a pair of donor and acceptor located in hot spots of metallic nanoparticle trimer as shown in the inset of Fig. 2(a), where the solid and dashed lines correspond to the cases with and without coherence, respectively. The radius of nanoparticles is taken as R=10 nm, and separation distance is d=1 nm. The dielectric function is taken as Drude model $\sigma(\omega) = i\varepsilon_0\omega_P^2/(\omega + i\eta)$ with plasma frequency $\omega_P$=9.01 eV, and loss rate $\eta$=0.048 eV[42]. The ambient media is water, $\varepsilon_0$=1.77. Comparing the two lines in Fig. 2(a), we find that the energy transfer efficiency has two significant changes when considering the coherent coupling. Not only the frequency shift happens, the peak value also nearly doubles. To analyze the reason of this phenomenon, in Fig. 2(b) and (c) we plot three parameters: coherent coupling $g_{12}/\gamma_{11}$, incoherent coupling $\gamma_{12}/\gamma_{11}$ and donor decay rate $\gamma_{11}$ as a function of the frequency.

According to the theoretical analysis in Sec. II, we know that there are two kinds of couplings between two molecules, coherent and incoherent coupling, in which incoherent coupling is resonant coupling. From Fig. 2(a), (b) and (c) we can see that, the peaks in three curves: energy transfer efficiency without coherence, incoherent coupling and donor decay rate, correspond to each other very well. That's to say, when we only consider incoherent coupling, the resonant coupling between molecules mediated by resonant field determines energy transfer process, in this case the energy transfer is the classical Förster resonance energy transfer[41]. While the case becomes very different when the coherent coupling is contained. After considering the coherence, the peaks for the energy transfer efficiency correspond to the peaks of coherent coupling strength $g_{12}/\gamma_{11}$, that's to say, the coherent coupling plays the major role in the increase and shift of the energy transfer efficiency. As described in Sec. II, coherent coupling between the two molecules is off-resonant coupling, it can shift the transition frequencies of molecules. This can explain why the frequencies of the peaks of coherent energy transfer efficiency deviate from resonant frequencies. According to Eq. (8), the improvement of the coherent energy transfer efficiency can be explained by using the population of the acceptor $P_A(t)$.

In Fig. 3, we plot the time evolution of the populations for the donor and acceptor when they are inserted in the gaps of nanoparticle trimer as shown in the inset of Fig. 2(a). The black dashed and red solid lines represent the populations of the donor and acceptor $P_D(t)$ and $P_A(t)$, respectively. Figure 3(a) represents the case that the coherent coupling $g_{12}/\gamma_{11}$=0 and incoherent coupling $\gamma_{12}/\gamma_{11}$ are in the peak value, and Fig. 3(b) is the case that $\gamma_{12}$=0 and $g_{12}/\gamma_{11}$ are in the peak value. The parameters of the system are taken the same as Fig. 2(a). It can be seen from Fig. 3(a) that, when coherent coupling $g_{12}/\gamma_{11}$=0, incoherent coupling can reach 0.97, which is very close to the theoretical maximum. However the population of the acceptor arrives at 0.2 and drops to zero gradually. By contrast, in Fig. 3(b), the population of the acceptor can reach 0.5 at the beginning and oscillates for a while due to the large coherent coupling, which makes the excitation exists in the acceptor for a longer time. This can improve the probability of charge separation and the energy transfer efficiency.

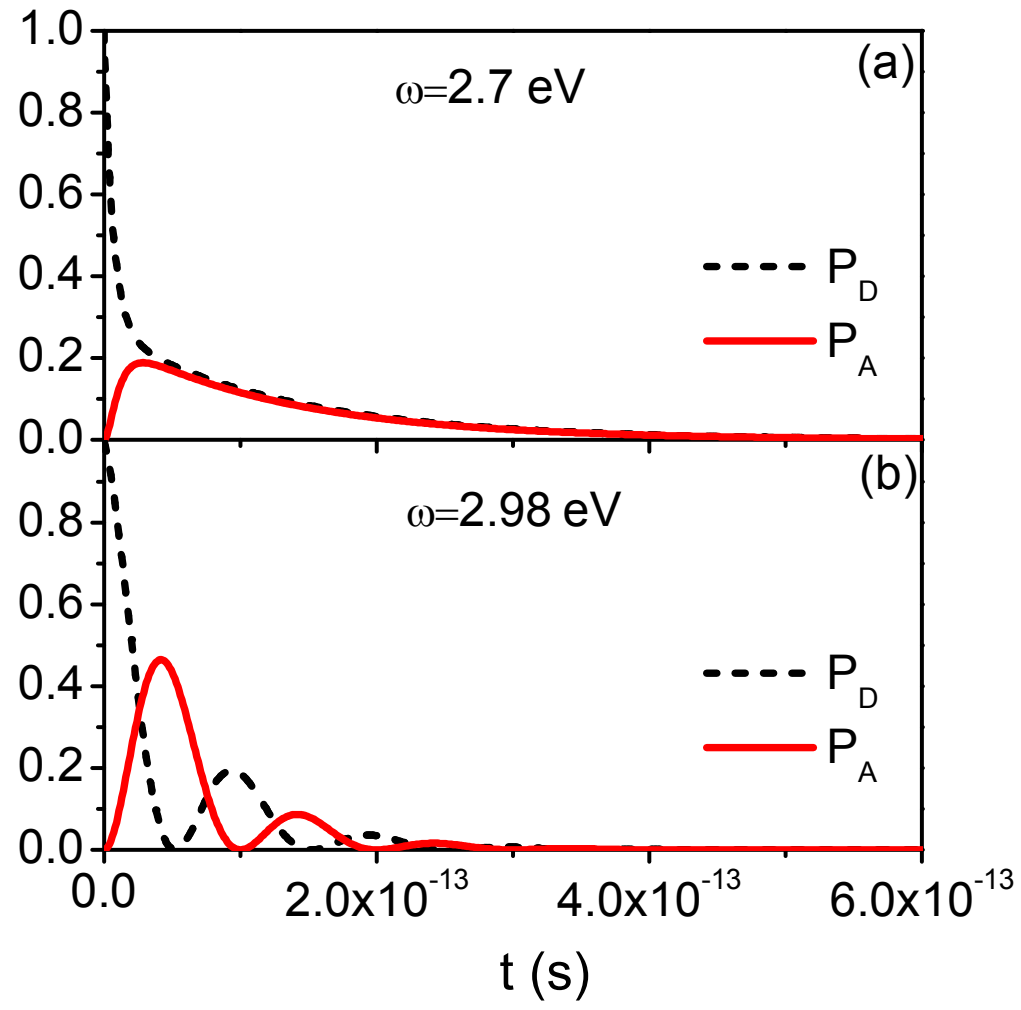

FIG. 3. Populations of donor and acceptor $P_D$ and $P_A$ when the donor and acceptor inserted in the hot spots of the nanoparticle trimer (see top inset), and the excitation is in donor initially. (a) $\omega$=2.7 eV, $\gamma$=$1.26\times10^{14}$, $g_{12}$=0, $\gamma_{12}$=0.97, d=e|1Å|. (b) $\omega$=2.98 eV, $\gamma$=$1.28\times10^{13}$, $g_{12}$=-2.46 $\gamma_{12}$=0, d=e|1Å|.

As comparison, in Fig. 2(d)-(f) we study the efficiency of the energy transfer between the two molecules with only single nanoparticle, as shown in the inset of Fig. 2(d). The parameters of the two molecules and center sphere are the same with the trimer case. Comparing the two lines in Fig. 2(d) we find that, for the energy transfer efficiency, the two cases with and without coherence have small difference in peak positions and values, and the efficiency is no more than 10%, much less than 40% in the trimer system. These phenomena can be explained from the coherent coupling strength presented in Fig. 2(e). In the single sphere case, the coherent coupling strength is less than 0.33, much smaller than 2.5 in the trimer system as shown in Fig. 2(b). This small coherent coupling leads to small energy transfer efficiency, and the two cases with and without coherence have small difference.

In addition, when there is only one nanoparticle in the system (Fig. 2(d)-(f)), we find that the peaks of the energy transfer efficiency and coherent coupling are in the frequency range of 4.0~5.3 eV. In this frequency range, the spontaneous decay rate has some peaks in the trimer system, which are similar to the single sphere case (Fig. 2(f)). These peaks originate from resonance modes, but they cannot arise large coupling and energy transfer efficiency. We call them single scattering resonance peaks, which are caused by single sphere scattering, thus can be observed in both single and trimer systems. By contrast, in trimer system, large energy transfer efficiency and strong coherence appear in the frequency range of 2.0~4.0 eV, and there are some peaks, which originate from coupling resonance among multispheres, and cannot be observed in the single sphere system. These peaks are called coupling resonance peaks, and this frequency range is called coupling resonance region. Due to the contribution of the large coherent coupling strength in the coupling resonance region, the large energy transfer efficiency in the trimer system can be achieved.

In fact, the efficient energy transfer also depends on the separation distance d between the spheres in the trimer system. Figure 2 only shows the case that d=1 nm, while in Fig. 4 we plot the dependence of energy transfer efficiency $\eta$ and corresponding parameters on d. In Fig. 4(a)-(d) we present $\eta$, $g_{12}/\gamma_{11}$, $\gamma_{12}/\gamma_{11}$ and $\gamma_{11}$ as a function of transition frequency, in which black solid, green dashed, blue dash dotted and red short dashed lines correspond to the results of d=1 nm, 2 nm, 4 nm and 8 nm. The pink short dotted line in Fig. 4(a) represents the energy transfer efficiency when d=12 nm. The radius of the spheres is fixed at R=10 nm. We find that, when the separation distance is very large, like d=12 nm, the energy transfer efficiency is smaller than 10%. The peaks of the efficiency span in the frequency range of 4.0~5.0 eV, which is very similar to the single sphere case as shown in Fig. 2(d). With the decrease of d, the energy transfer efficiency increases rapidly. When d decreases to 2~4 nm, the efficiency can reach 50%. While, when d is rather small, like 1 nm, the efficiency decreases to about 40%. These phenomena can be explained by Fig. 4(b)-(d). When d increases from 1 nm to 4 nm, for coherent coupling $g_{12}/\gamma_{11}$, the maximum of $g_{12}/\gamma_{11}$ shifts but has no apparent decrease in value. However when d=1 nm, donor decay rate is very large, that means the dissipation is large, thus energy transfer efficiency decreases. When d increases to 8 nm, there is an apparent decrease in coherent coupling, marked with triangles as shown in Fig. 4(b), this results in the decrease of energy transfer efficiency. For the incoherent coupling, from d=1 nm to d=8 nm, there is only frequency shift but no apparent decrease, see Fig. 4(c). Our results reveal that the energy transfer efficiency is mainly determined by the coherent coupling strength between the molecules. The stronger the coherent coupling strength, the larger the energy transfer efficiency. Moreover, the frequencies of coupling resonance peaks have compact relation with d, and single scattering resonance peaks have no relation with d, this could be seen in Fig. 4(d). With the increase of d, the coupling resonance peaks become closer to the single scattering resonance peaks. When d is very large (d>8 nm), the coupling resonance among the three namoparticles is very weak, only single scattering peaks appear.

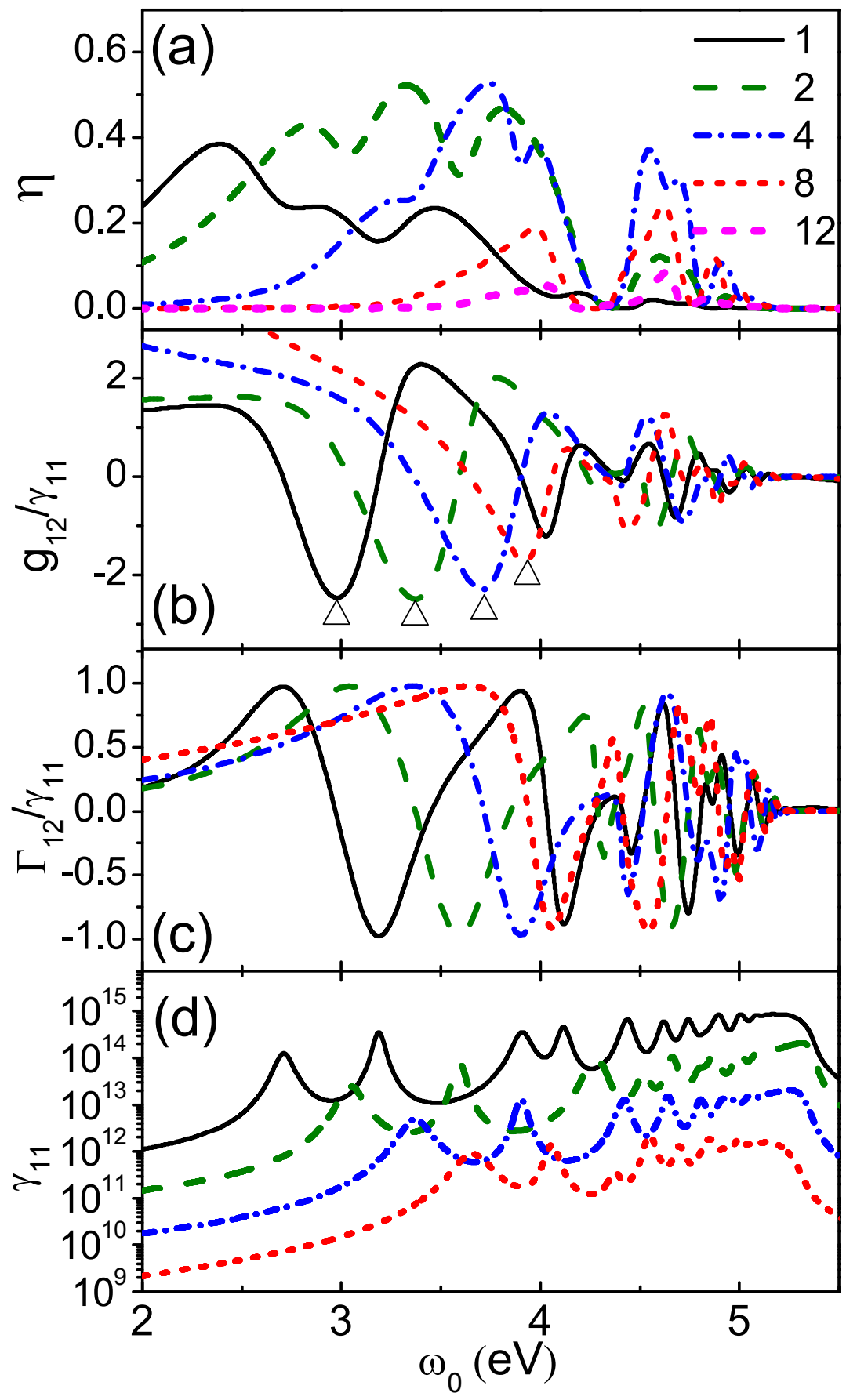

FIG. 4. The energy transfer efficiency and coupling parameters of a pair of donor and acceptor located in the hot spots of trimer as a function of transition frequency under different separation distance d. The unit in legend is nanometer.

In the following we calculate the energy transfer time $t_f$ according to Eq. (10). Figure 5 shows the calculated results in three different cases. The black solid line represents the trimer case, in which the coherence is considered, and the parameters of the system is the same to Fig. 2. The red dashed line is the case that $g_{12}/\gamma_{11}=0$ in the trimer system. The blue dotted line describes the energy transfer time in the single nanoparticle system. Firstly, it should be pointed out that energy transfer efficiency and time are the two key concepts in the coherent energy transfer theory, an efficient energy transfer should has large transfer efficiency and short transfer time. If transfer efficiency is almost zero, and then a short transfer time makes no sense. For example, we can see from the black solid line in Fig. 4(a) that, when d=1 nm, out of the frequency range of 2.0~4.0 eV, the energy transfer efficiency is almost zero. That means transfer time is meaningless out of this range. In this frequency range, the energy transfer time is minimum, only 60 fs, which is close to the energy transfer time in the photosynthesis[2], and much less than thousands of femtoseconds in the single sphere system. Moreover, after considering the coherence, the energy transfer time is less than the case without coherence, that means the coherent coupling can reduce energy transfer time. Figure 5(b) shows the energy transfer time under different separation distance d, in which black solid, magenta dashed, violet dotted and green dot dashed lines correspond to the results of d=1 nm, 4 nm, 8 nm and 12 nm. We can see that, with the increase of d, in the coupling resonance region (2.0~4.0 eV), $t_f$ has reached hundreds of femtoseconds when d=4 nm. Only for the small distance d, the energy transfer time can be very small.

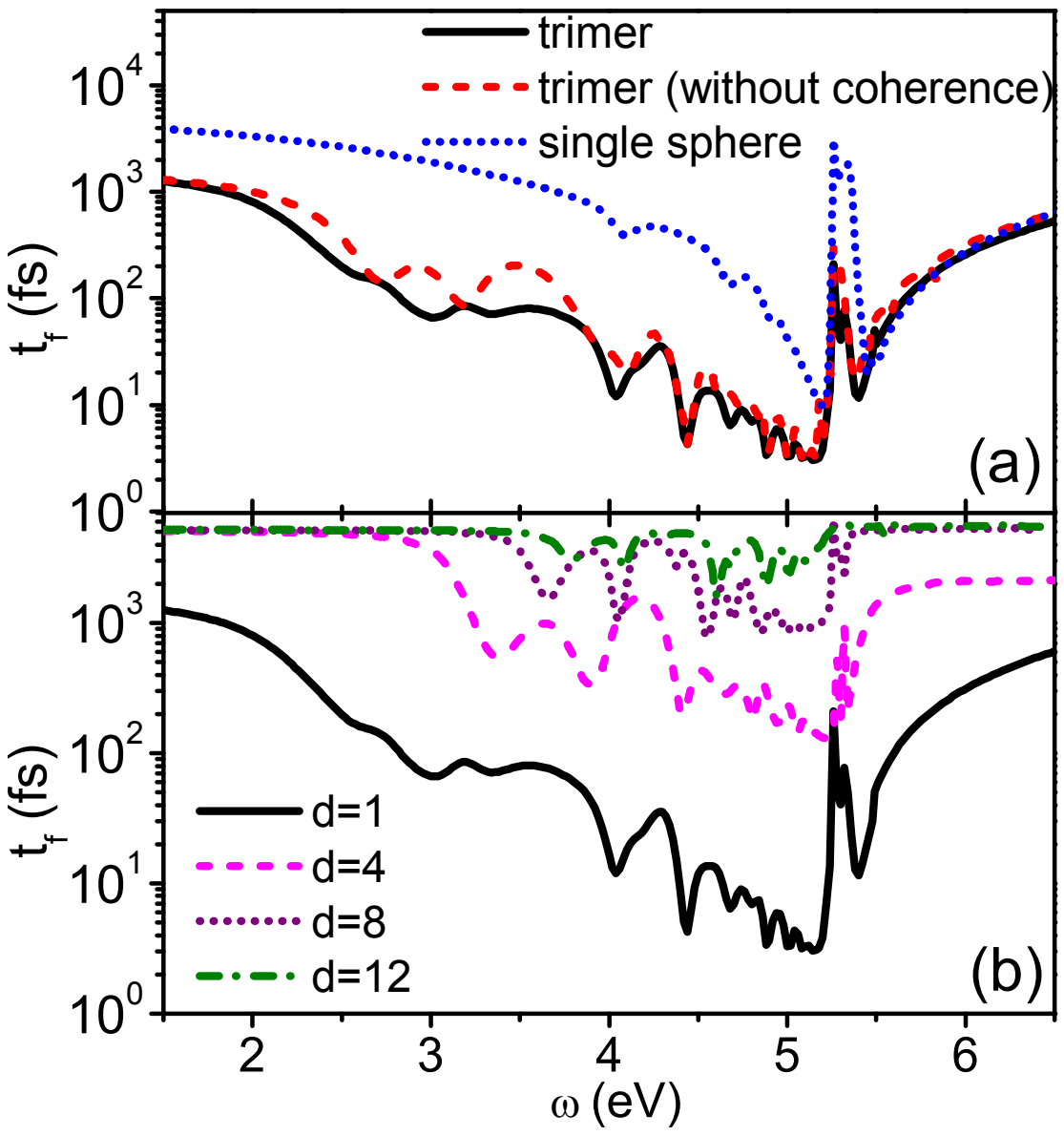

FIG. 5. (a) Energy transfer time of a pair of donor and acceptor located in hot spots of trimer (black solid line), as comparison, single sphere case (blue dotted line) and no coherence case (red dashed line) are also presented. (b) The dependence of energy transfer time $t_f$ on separation distance d.

## V. THE INFLUENCE OF NONLOCAL EFFECT ON COHERENT ENERGY TRANSFER EFFICIENCY AND TIME

All the calculations and discussions in Sec. III only consider the case that R=10 nm. In fact, the coherent coupling strength and coherent energy transfer efficiency also depend on the sizes of the spheres. Many recent investigations have shown that when the nanoparticle radius is less than 10 nm or the separation distance between two spheres is smaller than 1 nm, the effects of quantum size or nonlocal effect on the plasmon resonance will play an important role[42-45]. In the following, we will employ a hydrodynamic model to study the effect of nonlocal response on the coupling parameters and coherent energy transfer. For the single sphere, the field distribution and Green tensor with the nonlocal effect can be solved exactly according to Ref. 42. For the trimer system discussed above, we should include the nonlocal effect of every sphere, and the total effects can be considered exactly by using the multi-scattering T-matric method as shown in Appendix A. Here the Fermi velocity for the metal Ag is taken as $V_F = 1.39\times10^6$ m/s[42], and the dielectric function of Ag spheres are the same to Fig. 2.

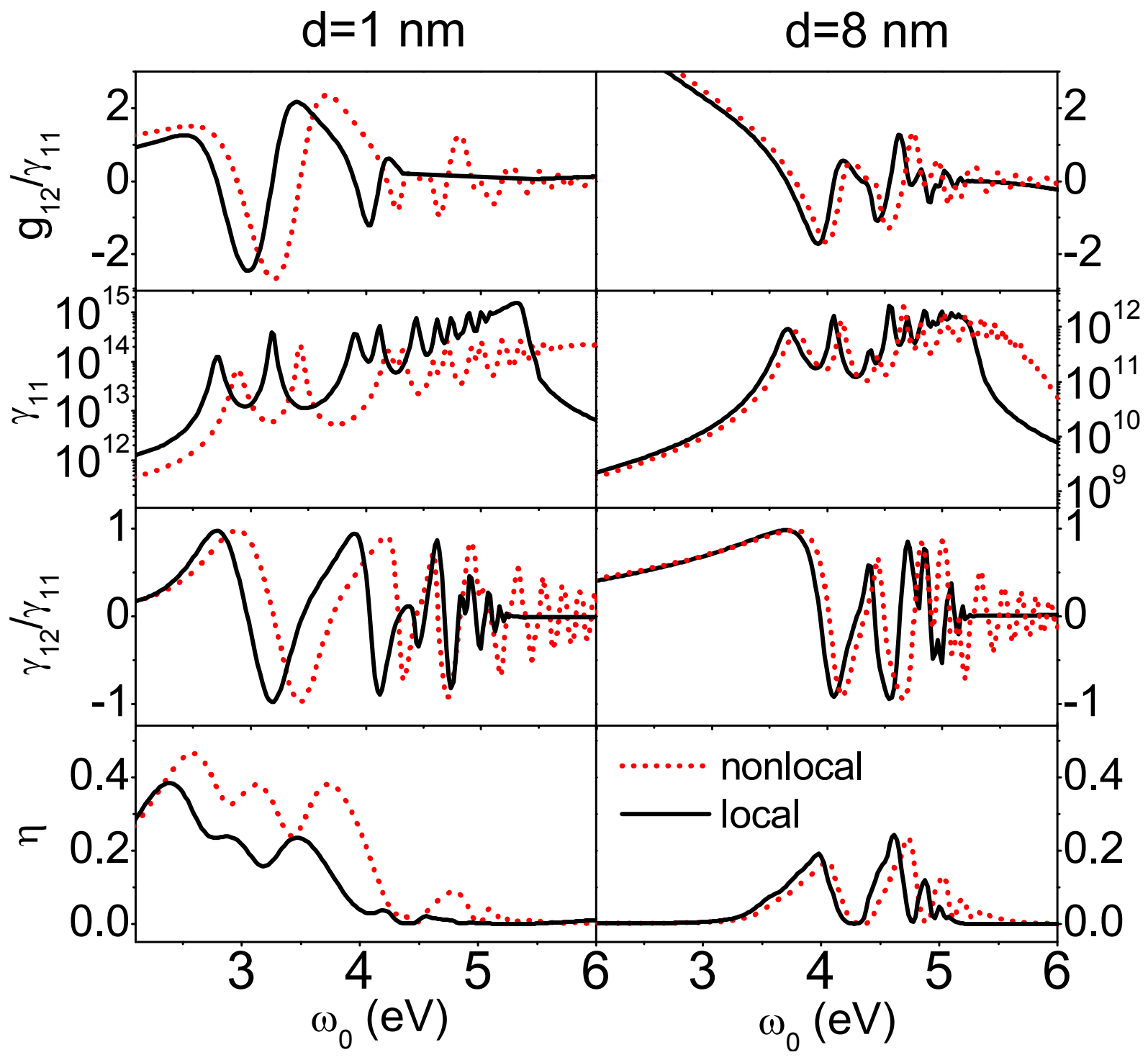

FIG. 6. The influence of nonlocal effect on coherent term $g_{12}/\gamma_{11}$, donor decay rate $\gamma_{11}$, incoherent term $\gamma_{12}/\gamma_{11}$ and coherent energy transfer efficiency $\eta$ at different separation distances. The radii of the sphere is fixed at R=10 nm. For the left column d=1 nm, right column d=8 nm. The black lines correspond to the results that without the nonlocal effect, the red lines correspond to those with the nonlocal effect.

Figure 6 displays the coherent coupling $g_{12}/\gamma_{11}$, incoherent term $\gamma_{12}/\gamma_{11}$ and donor decay rate $\gamma_{11}$ as a function of frequency under different separation distance d, where the radii of the spheres is fixed at R=10 nm. The black solid and red dashed lines correspond to the calculated results with and without nonlocal effect, respectively. The left column is for d=1 nm, and right column describes the case that d=8 nm. We find that, when the nonlocal effect is introduced, for d=1 nm, all the coherent coupling, incoherent coupling and donor decay rate have blue shifts, thus the energy transfer efficiency also has a blue shift. Due to the nonlocal effect, the local density of states in the positions of molecules decreases, and thus the spontaneous decay rate decreases. Under such circumstance, there is no apparent decrease in the coupling strength, in this way, the energy transfer efficiency will increase compared with the energy transfer efficiency without nonlocal effect. When d increases to 8 nm, the changes of all the parameters and energy transfer efficiency induced by the nonlocal effect are very small, and thus can be neglected. In addition we can see that, no matter d=1 nm or d=8 nm, when the frequency is larger than 5.0 eV, the donor decay rate and incoherent coupling have some new resonance peaks when considering the nonlocal effect. These peaks are caused by plasmons from the bulk material, and they appear no matter the size of the gap. Such peaks are not related to the strength of coherent coupling and therefore cannot affect the energy transfer efficiency.

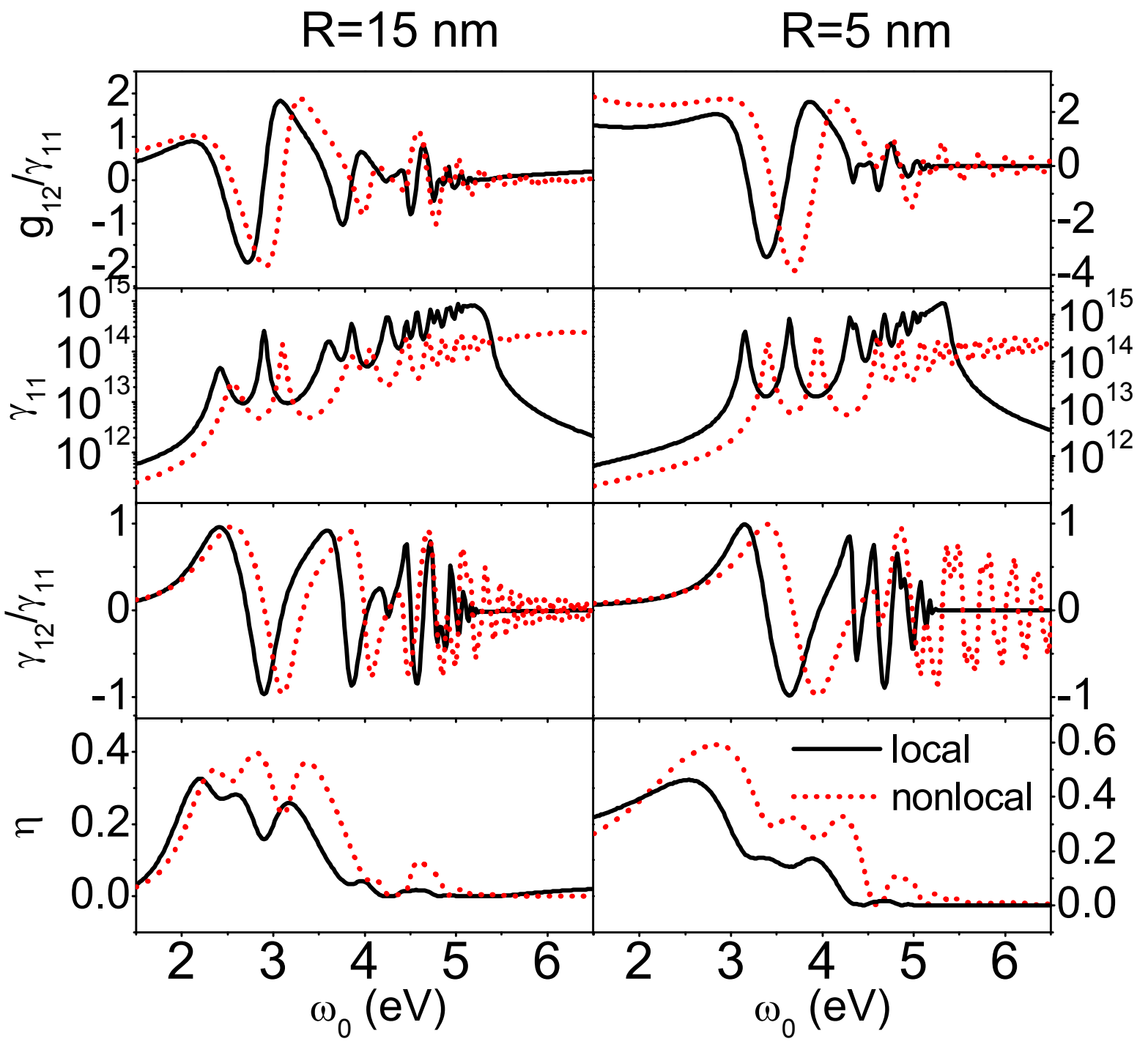

FIG. 7. The influence of nonlocal effect on coherent term $g_{12}/\gamma_{11}$, donor decay rate $\gamma_{11}$, incoherent term $\gamma_{12}/\gamma_{11}$ and coherent energy transfer efficiency $\eta$ with different sizes of spheres. The separation distances between the spheres are fixed at d=1 nm. For the left column R=15 nm, right column d=5 nm. The black lines correspond to the results without the nonlocal effect, the red lines to those with the nonlocal effect.

Figure 7 displays the coherent coupling $g_{12}/\gamma_{11}$, incoherent term $\gamma_{12}/\gamma_{11}$ and donor decay rate $\gamma_{11}$ as a function of the frequency under different sphere radii, where the separation distance d is fixed at d=1 nm. The black solid and red dashed lines correspond to the calculated results with and without the nonlocal effect. The left column is the result of R=15 nm, and the right column is for R=5 nm. It is clear that, no matter R=15 nm or 5 nm, due to the small separation distance, the nonlocal effect can cause blue shifts, and such shifts are more significantly than the case with d=5 nm. Based on the analysis above, both the separation distance d and the dimension of the spheres R can affect the energy transfer efficiency. The large energy transfer efficiency can be observed even though considering the nonlocal effect. When the separation distance is very large, the nonlocal effect can be ignored.

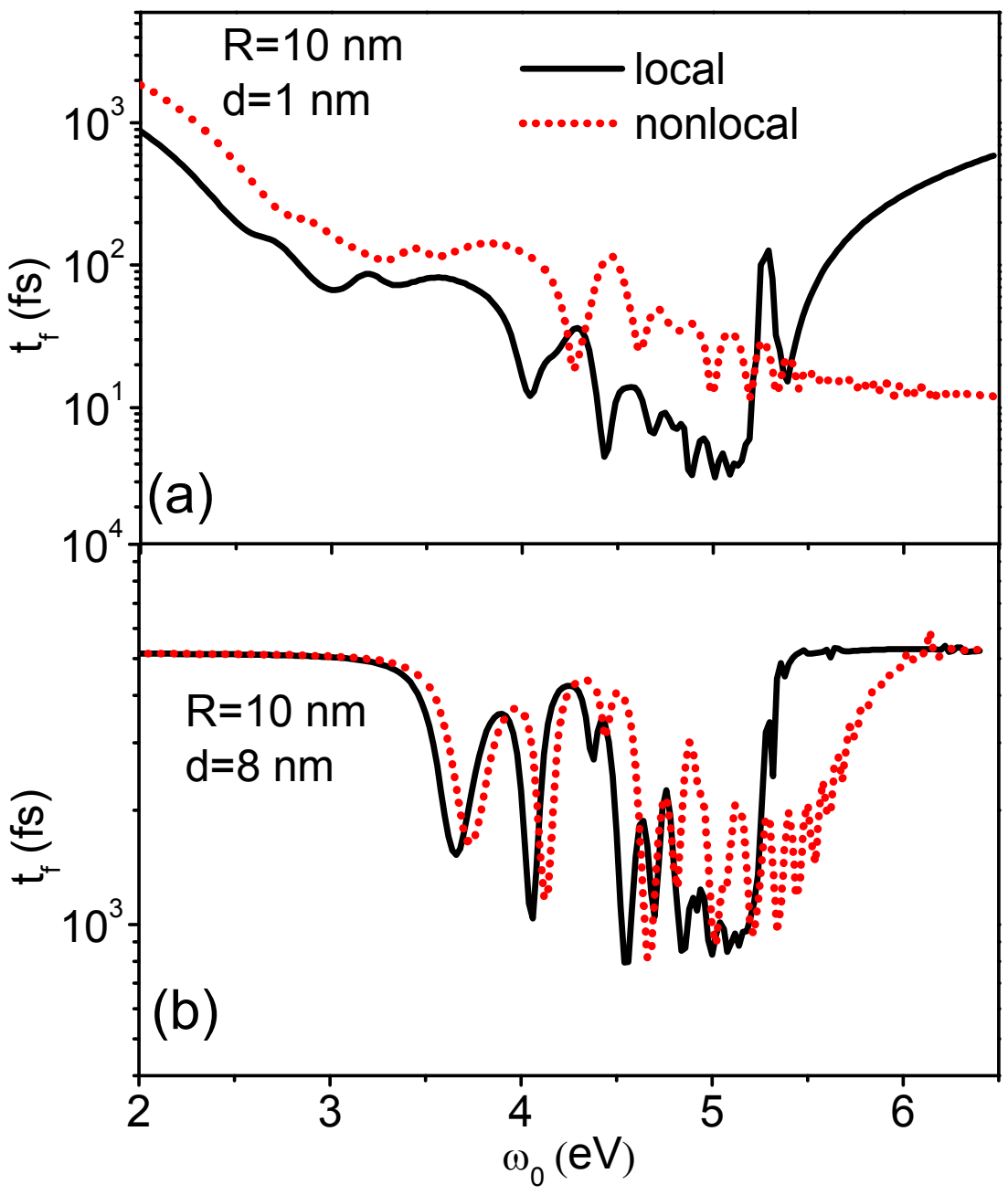

FIG. 8. The influence of nonlocal effect on the energy transfer time $t_f$ under different separation distance. The radii of the spheres is fixed at R=10 nm. (a) d=1 nm, (b) d=8 nm. The black lines correspond to the results without the nonlocal effect, the red lines to those with the nonlocal effect.

In addition to the energy transfer efficiency, the energy transfer time is also affected by the nonlocal effect. Figure 8 shows the energy transfer time $t_f$ as a function of the frequency when d=1 nm (Fig. 8(a)) and d=8 nm (Fig. 8(b)), where the radii of the sphere is fixed at R=10 nm. The black solid and red dashed lines correspond to the calculated results with and without the nonlocal effect. It can be seen that, when the nonlocal effect is introduced in the case that d=1 nm, $t_f$ shifts and increases. For instance, without the nonlocal effect, there is an minimum at $\omega$=4.0 eV with $t_f$=10 fs; in comparison, when considering the nonlocal effect, this minimum shifts to $\omega$=4.26 eV and $t_f$ increases to 20 fs. For another minimum of $t_f$, it changes from 4 fs at 4.43 eV to 23 fs at 4.62 eV. It can be seen that, although the shift is obvious, the increase of $t_f$ is not too large. When d=8 nm, both the shift and increase are very small, thus the nonlocal effect can be ignored.

Based on the discussions above we find that, when the radii of spheres is smaller than 10 nm or the separation distance is very small like 1 nm, the nonlocal effect will result in two obvious changes in the energy transfer. One is the blue shift in the frequency, the other is the small increase of coherent energy transfer efficiency and time. But the phenomena for ultrafast and efficient coherent energy transfers induced by the resonance coupling still exist. When the separation distance is as large as several nanometers, the nonlocal effect can be ignored.

## VI. CONCLUSIONS

The ultrafast energy transfer with high efficiency has been achieved in the plasmonic hot spot system with the introduction of the quantum coherence. In our study, we develop a theoretical method that combines the master equation approach and the multi-scattering T-matrix approach to investigate the coherent energy transfer. We find that with the appropriate frequencies and the system parameters (radii of the nanoparticle, separation distance etc.), the coherent coupling strength between the donor and the acceptor in plasmonic hot spot system becomes very strong due to the emergence of the quantum coherence. This strong coherent coupling leads to the high-efficiency and fast energy transfer. Moreover, we have investigated the influence of the nonlocal effect on the energy transfer efficiency and time. The high-efficiency and fast energy transfer can also be reached when the nonlocal effect is contained. Our results reveal the importance of the quantum coherence in the energy transfer for the strong coupling plasmonic system and can be applied into the fabrication of the efficient light-harvesting devices.

## ACKNOWLEDGMENTS

This work was supported by the National Key Basic Research Special Foundation of China under Grant 2013CB632704 and the National Natural Science Foundation of China (11604014 and 61421001).

## APPENDIX A. THE CALCULATION OF GREEN TENSOR BY T-MATRIX METHOD

The decay rates $\gamma_{ii(jj)}$, incoherent coupling $\gamma_{12}$ and coherent coupling $g_{12}$ can be obtained from the calculations of Green tensor. The classical Green tensors in Eqs. (2) and (3) can be calculated by the T-matrix method. The dyadic Green's function $\vec{G}(\vec{r}_i, \vec{r}_j, \omega)$ represents the electric field at position $\vec{r}_i$ in the nanoparticle cluster excited by a unit dipole at position $\vec{r}_j$. In the framework of the T-matrix approach[46-48], the incident and scattered fields are expanded in vector spherical functions (VSFs):

$$E_{inc}(\boldsymbol{r}-\boldsymbol{R}_i)=\sum_{v=1}^{\infty}a_v^i\boldsymbol{M}_v^1\left(k\left(\boldsymbol{r}-\boldsymbol{R}_i\right)\right)+b_v^i\boldsymbol{N}_v^1\left(k\left(\boldsymbol{r}-\boldsymbol{R}_i\right)\right), \quad \text{(A1)}$$

$$E_s^i(\boldsymbol{r}-\boldsymbol{R}_i)=\sum_{v=1}^{\infty}f_v^i\boldsymbol{M}_v^3\left(k(\boldsymbol{r}-\boldsymbol{R}_i)\right)+g_v^i\boldsymbol{N}_v^3\left(k(\boldsymbol{r}-\boldsymbol{R}_i)\right) \quad |r_i|>\Re_i, \quad \text{(A2)}$$

where $\boldsymbol{M}_v^1$ , $\boldsymbol{N}_v^1$ , $\boldsymbol{M}_v^3$ and $\boldsymbol{N}_v^3$ are the well-known VSFs, and $\boldsymbol{r}_i$ is a position vector in the coordinate of the $ith$ particle. $\Re_i$ is radius of the smallest sphere circumscribing the $ith$ object. $a_v^i$ , $b_v^i$ , $f_v^i$ and $g_v^i$ are the expansion coefficients, which can be readily known as soon as the form of the incident wave is given. $v$ stands for (m, n) which are the indices of spherical harmonic functions. At the same time the internal field of the $ith$ nanoparticle is written as,

$$E_{int}^i(\boldsymbol{r}-\boldsymbol{R}_i)=\sum_{v=1}^{\infty}c_v^i\boldsymbol{M}_v^1\left(k_i(\boldsymbol{r}-\boldsymbol{R}_i)\right)+d_v^i\boldsymbol{N}_v^1\left(k_i(\boldsymbol{r}-\boldsymbol{R}_i)\right). \quad \text{(A3)}$$

According to the T-matrix method[46-48], $c_v^i$ and $d_v^i$ are related to $a_v^i$ and $b_v^i$ by the following matrix equation:

$$\begin{bmatrix}\boldsymbol{Q}_i^{11} & \boldsymbol{Q}_i^{12}\\ \boldsymbol{Q}_i^{21} & \boldsymbol{Q}_i^{22}\end{bmatrix}\begin{bmatrix}\boldsymbol{c}^i\\ \boldsymbol{d}^i\end{bmatrix}=\sum_{j=1}^{N(j\neq i)}\begin{bmatrix}\boldsymbol{T}_{ij}^{11} & \boldsymbol{T}_{ij}^{12}\\ \boldsymbol{T}_{ij}^{21} & \boldsymbol{T}_{ij}^{22}\end{bmatrix}\begin{bmatrix}Rg\boldsymbol{Q}_j^{11} & Rg\boldsymbol{Q}_j^{12}\\ Rg\boldsymbol{Q}_j^{21} & Rg\boldsymbol{Q}_j^{22}\end{bmatrix}\begin{bmatrix}\boldsymbol{c}^j\\ \boldsymbol{d}^j\end{bmatrix}+\begin{bmatrix}\boldsymbol{a}^i\\ \boldsymbol{b}^i\end{bmatrix}, \quad \text{(A4)}$$

where $\boldsymbol{Q}_i^{pq}$ and $Rg\boldsymbol{Q}_j^{pq}$ are the T-matrix blocks for the $ith$ and $jth$ particles, and $\boldsymbol{T}_{ij}^{pq}$ is block of the transition matrix between the $ith$ particle and the $jth$ particle[46-48]. By solving these equations, expansion coefficients of the inner field for each object can be known. And also according to the equation:

$$\begin{bmatrix}\boldsymbol{f}^i\\ \boldsymbol{g}^i\end{bmatrix}=\begin{bmatrix}Rg\boldsymbol{Q}^{11} & Rg\boldsymbol{Q}^{12}\\ Rg\boldsymbol{Q}^{21} & Rg\boldsymbol{Q}^{22}\end{bmatrix}\begin{bmatrix}\boldsymbol{c}^i\\ \boldsymbol{d}^i\end{bmatrix}, \quad \text{(A5)}$$

the scattering expansion coefficients $f_v^i$ and $g_v^i$ of each particle can be easily calculated. The field outside the circumscribing spheres then can be obtained using the following equation:

$$\boldsymbol{E}_{ext}=\boldsymbol{E}_{inc}+\sum_{i=1}^{N}\boldsymbol{E}_s^i . \quad \text{(A6)}$$

About the calculation of the external scattering field $\boldsymbol{E}_d^{\kappa}$ induced by a dipole with a momentum of $\mu_{12}^{\kappa}$, we take the exciting source as a dipole $\boldsymbol{\mu}_{12}^{\kappa}$ located in $\boldsymbol{r}_{\kappa}$ , the incident wave can be

expressed as

$$\boldsymbol{E}_{inc}=\frac{-\nabla\times\nabla\times\boldsymbol{A}_p^{(\kappa)}}{i\omega\mu_0\varepsilon_{med}\varepsilon_0},\tag{A7}$$

where

$$\boldsymbol{A}_p^{(\kappa)}=-i\omega\mu_0\frac{e^{ik|\boldsymbol{r}-\boldsymbol{r}_\kappa|}}{4\pi|\boldsymbol{r}-\boldsymbol{r}_\kappa|}\boldsymbol{\mu}_{12}^{\kappa}.\tag{A8}$$

Expanding Eq. (A7) to the same form of Eq. (A1), from Eqs. (A2-A6) we can obtain the external scattering field $\boldsymbol{E}_d^{\kappa}=\sum_{i=1}^{N}\boldsymbol{E}_s^i$ caused by the dipole. Then, the dyadic Green's function $\vec{G}(\vec{r}_i,\vec{r}_j,\omega)$ can be obtained

$$\vec{n}_i\cdot\vec{G}\left(\vec{r}_i,\vec{r}_j;\omega\right)\cdot\vec{n}_j=-\vec{n}_i\cdot\boldsymbol{E}_d^{\kappa},\tag{A9}$$

where $\vec{n}_i$ and $\vec{n}_j$ are the unit vectors of the field and source dipole moments, respectively.

## APPENDIX B. COHERENT COUPLING TERM

In this part, we give a detailed derivation for Eq. (6). For the coherent term appeared in Eq. (3), we have to deal with the principle integral in our multi-particle system. Similar to Ref. 40, the tough principle integral can be avoided by using the Kramers-Kronig relation and contour integral. For the complex-valued function $f(\omega)$, the K-K relation is

$$\operatorname{Re}f(\omega_A)=\frac{1}{\pi}\mathcal{P}\int_{-\infty}^{\infty}\frac{\operatorname{Im}f(\omega)}{\omega-\omega_A}\mathrm{d}\omega,\tag{B1}$$

$$\operatorname{Im}f(\omega_A)=-\frac{1}{\pi}\mathcal{P}\int_{-\infty}^{\infty}\frac{\operatorname{Re}f(\omega)}{\omega-\omega_A}\mathrm{d}\omega,\tag{B2}$$

where the integral interval is from $-\infty$ to $\infty$. By using the relations $\vec{G}(-\omega^*)=\vec{G}^*(\omega)$ and $\operatorname{Im}\vec{G}^*(\omega)=-\operatorname{Im}\vec{G}(\omega)$, and let $f(\omega)=(\omega^2/c^2)\vec{G}(\omega)$, we can obtain the relation

$$\frac{\omega_A^2}{c^2}\operatorname{Re}\vec{G}(\omega_0)=\frac{2}{\pi}\mathcal{P}\int_0^{\infty}\frac{\omega^2}{c^2}\frac{\omega\operatorname{Im}\vec{G}(\omega)}{\omega^2-\omega_0^2}\mathrm{d}\omega.\tag{B3}$$

Equation (6) arrives at

$$g_{ij}=\frac{1}{\pi\varepsilon_0\hbar}\mathcal{P}\int_0^{\infty}\frac{\omega^2\operatorname{Im}\left[\vec{\mu}_i^*\cdot\vec{G}(\vec{r}_i,\vec{r}_j,\omega)\cdot\vec{\mu}_j\right]}{c^2\left(\omega^2-\omega_0^2\right)}\left(\omega+\omega_0\right)d\omega.\tag{B4}$$

Substituting Eq. (B3) to Eq. (B4), we obtain

$$g_{ij} = \frac{1}{\pi\varepsilon_0\hbar}\left\{\frac{\pi}{2}\frac{\omega_A^2}{c^2}\mathrm{Re}\left[\vec{\mu}_i^* \cdot \vec{G}(\vec{r}_i,\vec{r}_j,\omega)\cdot\vec{\mu}_j\right] + \mathcal{P}\int_0^\infty \frac{\omega^2}{c^2}\frac{\mathrm{Im}\left[\vec{\mu}_i^* \cdot \vec{G}(\vec{r}_i,\vec{r}_j,\omega)\cdot\vec{\mu}_j\right]}{\omega^2-\omega_0^2}\omega_0 d\omega\right\}. \quad \text{(B5)}$$

Equation (B5) still contains principle integral, in the following we will deal with the second term

$$\mathrm{Im}\left\{\mathcal{P}\int_0^\infty \frac{\omega^2}{c^2}\left[\vec{\mu}_i^* \cdot \vec{G}(\vec{r}_i,\vec{r}_j,\omega)\cdot\vec{\mu}_j\right]\frac{\omega_0}{\omega^2-\omega_0^2}d\omega\right\}. \quad \text{(B6)}$$

Let

$$f(\omega) = \frac{\omega^2}{c^2}\left[\vec{\mu}_i^* \cdot \vec{G}(\vec{r}_i,\vec{r}_j,\omega)\cdot\vec{\mu}_j\right]\frac{\omega_0}{\omega^2-\omega_0^2}, \quad \text{(B7)}$$

the integral in Eq. (B6) turns to

$$\mathcal{P}\int_0^\infty f(\omega)d\omega. \quad \text{(B8)}$$

In complex plane showed in the figure below, the contour $l$ is taken as the anticlockwise quarter circle in the first quadrant.

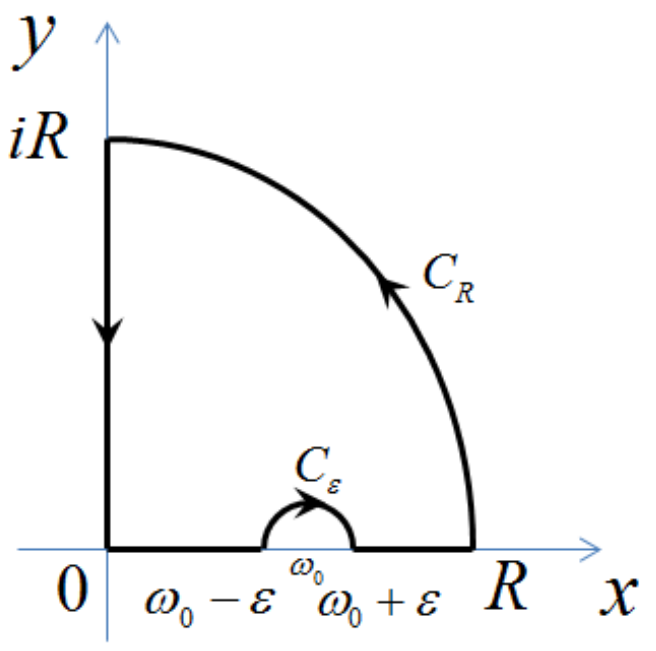

FIG. 9. The integral contour $l$ in complex plane, which is taken as the anticlockwise quarter circle in the first quadrant.

Assuming the integrand $f(z)$ only has one singularity in the positive real axis, the integral along contour $l$ can be obtained by avoiding the singularity $z=\omega_0$

$$\begin{aligned}\oint_l f(z)\,\mathrm{d}z &= \int_0^{\omega_0-\varepsilon} f(\omega)\,\mathrm{d}\omega + \int_{\omega_0+\varepsilon}^{R} f(\omega)\,\mathrm{d}\omega \\ &+ \int_{C_R} f(z)\,\mathrm{d}z + \int_{C_\varepsilon} f(z)\,\mathrm{d}z + \int_{iR}^{0} f(z)\,\mathrm{d}z\end{aligned}. \quad \text{(B8)}$$

According to Residue theorem, the left hand of Eq. (B8) is equal to zero for there is no singularity in the internal of the contour. While for the right hand, when $R\to\infty$ and $\varepsilon\to 0$, the sum of the first and second term is the principle integral Eq. (B8). The third term could be proven to zero by Jordan's lemma. For the fourth term, expand $f(z)$ for Laurent series at the neighborhood of $z=\omega_0$

$$f(z) = \frac{a_{-1}}{z-\omega_0} + \mathrm{P}(z-\omega_0), \quad \text{(B9)}$$

where $\mathrm{P}(z-\omega_0)$ is the analytic part of Laurent series, it is continuous and bounded on $C_\varepsilon$, therefore

$$\begin{aligned}\left|\int_{C_\varepsilon}\mathrm{P}(z-\omega_0)dz\right| &\le \max\left|\mathrm{P}(z-\omega_0)\right|\int_{C_\varepsilon}|dz| \\ &= \pi\varepsilon\cdot\max\left|\mathrm{P}(z-\omega_0)\right|\end{aligned}, \quad \text{(B10)}$$

and thus $\lim_{\varepsilon\to 0}\int_{C_\varepsilon}\mathrm{P}(z-\omega_0)dz=0$. For the first term of the right hand of Eq. (B9)

$$\begin{aligned}\int_{C_\varepsilon}\frac{a_{-1}}{z-\omega_0}dz &= \int_{C_\varepsilon}\frac{a_{-1}}{z-\omega_0}d(z-\omega_0) \\ &= \int_\pi^0\frac{a_{-1}}{\varepsilon e^{i\varphi}}d(\varepsilon e^{i\varphi}) = -\pi i a_{-1} = -\pi i\,\mathrm{Res}\,f(\omega_0)\end{aligned}, \quad \text{(B11)}$$

where the residue

$$\mathrm{Res}\,f(\omega_0)=\frac{1}{2}\frac{\omega_A^2}{c^2}\left[\vec{\mu}_i^*\cdot\vec{G}(\vec{r}_i,\vec{r}_j,\omega_0)\cdot\vec{\mu}_j\right]. \quad \text{(B12)}$$

Based on Eqs. (B9)-(B12), the fourth term turns to

$$\int_{C_\varepsilon}f(z)\,\mathrm{d}z=-\frac{\pi i}{2}\frac{\omega_0^2}{c^2}\left[\vec{\mu}_i^*\cdot\vec{G}(\vec{r}_i,\vec{r}_j,\omega_0)\cdot\vec{\mu}_j\right]. \quad \text{(B13)}$$

When $R\to\infty$, for the fifth term of the right hand of Eq. (B8), we have

$$\int_{iR}^0 f(z)\,\mathrm{d}z=-\int_0^{i\infty}f(\omega)\,\mathrm{d}\omega, \quad \text{(B14)}$$

let $\omega\to i\kappa$, Eq. (B14) turns to

$$\int_{iR}^0 f(z)\,\mathrm{d}z=-i\int_0^\infty\frac{\kappa^2}{c^2}\left[\vec{\mu}_i^*\cdot\vec{G}(\vec{r}_i,\vec{r}_j,i\kappa)\cdot\vec{\mu}_j\right]\frac{\omega_0}{\kappa^2+\omega_0^2}\mathrm{d}\kappa. \quad \text{(B15)}$$

Substituting equations (B13) and (B15) to Eq. (B8), we have

$$\begin{aligned}\mathcal{P}\int_0^\infty f(\omega)d\omega &= -\int_{C_\varepsilon}f(z)\,\mathrm{d}z-\int_{iR}^0 f(z)\,\mathrm{d}z \\ &= \frac{\pi i}{2}\frac{\omega_0^2}{c^2}\left[\vec{\mu}_i^*\cdot\vec{G}(\vec{r}_i,\vec{r}_j,\omega_0)\cdot\vec{\mu}_j\right]+i\int_0^\infty\frac{\kappa^2}{c^2}\left[\vec{\mu}_i^*\cdot\vec{G}(\vec{r}_i,\vec{r}_j,\omega_0)\cdot\vec{\mu}_j\right]\frac{\omega_0}{\kappa^2+\omega_0^2}\mathrm{d}\kappa\end{aligned}. \quad \text{(B16)}$$

Thus from Eq. (B5), we arrive at

$$\begin{aligned}g_{ij} &= \frac{\omega_A^2}{\varepsilon_0\hbar c^2}\mathrm{Re}\left[\vec{\mu}_i^*\cdot\vec{G}(\vec{r}_i,\vec{r}_j,\omega)\cdot\vec{\mu}_j\right] \\ &+\frac{1}{\pi\varepsilon_0\hbar}\int_0^\infty\mathrm{Re}\left[\vec{\mu}_i^*\cdot\vec{G}(\vec{r}_i,\vec{r}_j,i\kappa)\cdot\vec{\mu}_j\right]\frac{\omega_A}{\omega_A^2+\kappa^2}d\kappa\end{aligned}. \quad \text{(B17)}$$

These formulas are suitable for total Green tensor $\vec{G}$, vacuum Green tensor $\vec{G}^{vac}$, and scattering Green tensor $\vec{G}^s=\vec{G}-\vec{G}^{vac}$

$$\begin{aligned}g_{ij} &= \frac{\omega_A^2}{\varepsilon_0\hbar c^2}\mathrm{Re}\left[\vec{\mu}_i^*\cdot\vec{G}^s(\vec{r}_i,\vec{r}_j,\omega)\cdot\vec{\mu}_j\right] \\ &+\frac{1}{\pi\varepsilon_0\hbar}\int_0^\infty\mathrm{Re}\left[\vec{\mu}_i^*\cdot\vec{G}^s(\vec{r}_i,\vec{r}_j,i\kappa)\cdot\vec{\mu}_j\right]\frac{\omega_A}{\omega_A^2+\kappa^2}d\kappa\end{aligned}. \quad \text{(B18)}$$